\documentclass[%
reprint,
amsmath,amssymb,
aps,superscriptaddress
]{revtex4-2}

\usepackage{graphicx}
\usepackage{dcolumn}
\usepackage{bm}

\usepackage{epsfig}
\usepackage[english]{babel}
\usepackage{latexsym}
\usepackage{graphics}
\usepackage{subfigure}
\usepackage{graphicx}
\usepackage{amsmath}
\usepackage{hyperref}
\usepackage{amssymb}
\usepackage{color}

\begin{document}
	
	\preprint{APS/123-QED}
	
	\title{Identifying Mechanism of Energy-Resolved Attoclock}
	\author{J. N. Wu}
	\homepage{These authors contribute equally to this paper.}
	\affiliation{College of Physics and Information Technology and Quantum Materials and Devices Key Laboratory of Shaanxi Province's High Education Institution, Shaan'xi Normal University, Xi'an, China}
	\author{S. Q. Shen}
	\homepage{These authors contribute equally to this paper.}
	\affiliation{College of Physics and Information Technology and Quantum Materials and Devices Key Laboratory of Shaanxi Province's High Education Institution, Shaan'xi Normal University, Xi'an, China}
	\author{J. Y. Che}
	\affiliation{College of Physics, Henan Normal University, Xinxiang, China}
	\author{S. Wang}
	\email{phywangshang@163.com}
	\affiliation{College of Physics and Hebei Key Laboratory of Photophysics Research and Application, Physics Postdoctoral Research Station, Hebei Normal University, Shijiazhuang, China}
	\author{W. Y. Li}
	\email{liwy157@126.com}
	\affiliation{Hebei Key Laboratory of Optoelectronic Information and Geo-detection Technology, School of Mathematics and Science, Hebei GEO University, Shijiazhuang, China}
	\author{G. G. Xin}
	\affiliation{School of Physics, Northwest University, Xi'an, China}
	\author{Y. J. Chen}
	\email{chenyjhb@gmail.com}
	\affiliation{College of Physics and Information Technology and Quantum Materials and Devices Key Laboratory of Shaanxi Province's High Education Institution, Shaan'xi Normal University, Xi'an, China}
	
	\date{\today}
	
	\begin{abstract}
		We study above-threshold ionization (ATI) of atoms in strong elliptical laser fields numerically and analytically. Recent benchmark experiments for H showed that the attoclock offset angle related to each ATI ring increases remarkably with energy and this characteristic phenomenon can be attributed to the laser-induced nonadiabatic initial velocity and position of the electron at the tunnel exit [PRL127, 273201 (2021)]. However, the specific mechanism of how the nonadiabatic effects influence this angle remains unclear. Here, by using a strong-field model that analytically and quantitatively decouples complex nonadiabatic effects and Coulomb effects, the detailed mechanism can be clearly identified. We show that due to nonadiabatic effects, the angles  associated with lower (higher) energy rings are dominated by the main (minor) axis of the laser ellipse, jumping from $0^o$ to $90^o$. These field-related rigid effects are softened by Coulomb-induced exit velocity closely related to system symmetry, resulting in a significant but smooth increase in angle with energy.
	\end{abstract}
	
	\maketitle
	
	
	Tunneling ionization \cite{Keldysh1965, Faisal1973, Reiss1980, Ammosov1986, Tong2002} is the first step of many basic strong-field processes such as ATI \cite{Agostini1979, Yang1993, Paulus1994, Lewenstein1995, Becker2002} and high-harmonic generation (HHG) \cite{McPherson1987, Huillier1988, Corkum1993, Lewenstein1994, Krausz2009}. Attosecond angular streaking, also known as attoclock \cite{Eckle2008,Pfeiffer2012,Landsman2013,Camus2017, Quan2019,Sainadh2019,Han2021}, is an advanced technology for probing atomic and molecular tunneling dynamics using the photoelectron momentum distribution (PMD) generated by a strong elliptically-polarized laser (EPL) field. Generally, the most probable electron emission angle relative to the minor axis of the laser ellipse (also called the offset angle) is measured in attoclock as the characteristic quantity to analyze tunneling dynamics and validate theory models \cite{Boge2013,Torlina2015,Klaiber2015,Bray2018}. 
	
	Recent benchmark experiments on the simplest atom H showed that in strong short-wavelength EPL fields, well-resolved ATI rings can be observed in PMD, enabling attoclock to measure the offset angles associated with each ring \cite{Trabert2021}. It is found that the offset angle increases significantly with energy and this characteristic phenomenon can be attributed to the nonadiabatic effect associated with the laser-induced initial velocity and position of tunneling electrons at the tunnel exit. However, the detailed mechanism of this phenomenon, namely how the nonadiabatic effect affects the offset angle, is still unclear. The potential difficulties in unveiling this mechanism are twofold. Firstly, in experiments, energy-resolved attoclock can only be well achieved at larger Keldysh parameters \cite{Keldysh1965}, where in addition to tunneling, multiphoton transitions may also be involved, making the situation more complex. Secondly, theoretically, this phenomenon is generally analyzed using strong-field models, which consider Coulomb correction by numerically solving Newton equations, making it difficult to obtain a simple picture. 
	
	To overcome these difficulties, in this letter, firstly, we mainly analyze this phenomenon through numerical solution of time-dependent schrödinger equation (TDSE) at smaller Keldysh parameters, where the ATI rings can still be well resolved. Secondly, we use a recently developed strong-field model to analytically describe the Coulomb effect from the perspective of system symmetry, enabling analytical separation of nonadiabatic and Coulomb effects, making it easier to obtain a clear physical picture. The developed model gives the scaling law of this angle relative to energy, which is in good agreement with experimental results (see Fig. 1). We show that due to the nonadiabatic effect, when the angles related to low-energy rings are mainly determined by the main axis, the angles of high-energy rings are dominated by the minor axis, resulting in a sudden jump of this angle from $0^o$ to $90^o$. The Coulomb effect, characterized by a Coulomb-induced exit velocity, moderately regulates the field-dominating angle, resulting in a significant but gradual increase in this angle. This mechanism is somewhat different from the Coulomb-dominating mechanism in the general attoclock. The potential reason is associated with the geometry structure of the EPL field which does not matches the circular structure of the ATI ring (see Fig. 2(c)).  
	
	\textit{Theory model}. To analytically study the energy-resolved accoclock, we use a Coulomb-included strong-field model termed as tunneling-response-classic-motion (TRCM) model developed recently \cite{Che2021, Che2023, CheJ2023, Shen2024, Chen2025}. The model is based on strong-field approximation (SFA) \cite{Lewenstein1995} and electron-trajectory (or saddle-point) theory \cite{Becker2002}, but considers the near-nucleus Coulomb effect from the perspective of system symmetry \cite{Chen2025}. In SFA, the mapping relation between the drift momentum $\textbf{p}$ and the saddle-point time $t_s=t_0+it_x$ is $\textbf{p}=\textbf{v}(t_0)-\textbf{A}(t_0)$. Here, $\textbf{A}(t)$ is the vector potential of the laser electric field $\textbf{E}(t)$. The velocity $\textbf{v}(t_0)$, which is generally nonzero, is called the nonadiabatic initial velocity of the tunneling electron at the tunnel exit position $\textbf{r}_0(t_0)$ \cite{Yan2010}. In TRCM,  the Coulomb-modified drift momentum $\textbf{p}'$ can be written as
	\begin{equation}
		\textbf{p}'=\textbf{v}(t_0)+\textbf{v}_i-\textbf{A}(t_0)\approx\textbf{v}(t_0)-\textbf{A}(t_i),
	\end{equation}
	with $t_i=t_0+\tau$. Here, $\textbf{v}_i=-{v}_i\textbf{r}_0/{r}_0$ \cite{Goreslavski2004} with ${v}_{i}=k\sqrt{\left| V(\textbf{r}_0) \right|/{n_{f}}}$ which considers the Coulomb-related symmetry of the system in tunneling ionization, and $k=1.1$ being a small correction factor \cite{Chen2025}. The term $n_f$ is the dimension of the system studied, with $n_f=3$ ($n_f=2$) for three(two)-dimensional cases. The term $V(\textbf{r})$ is the Coulomb potential and $\tau\approx {v}_{i}/\left|\textbf{E}(t_0)\right|$ is the Coulomb-induced ionization time delay. The applicability of the expression for $\textbf{p}'$ has been validated in \cite{Che2021, Che2023, CheJ2023} through comparing with experiments.

	The amplitude $c(\textbf{p},t_0)$ for the SFA electron trajectory $(\textbf{p},t_0)$ can be written as $c(\textbf{p},t_0)\propto{\textbf{E}(t_s)\cdot \textbf{d}_i[\textbf{p}+\textbf{A}(t_s)]e^{-iS(\textbf{p},t_s)}}$, with the quasiclassical action $S(\textbf{p},t_s)=\int_{t_s}\lbrace[\textbf{p}+\textbf{A}(t')]^2/2+I_p\rbrace dt'=S_{re}(\textbf{p},t_s)+iS_{im}(\textbf{p},t_s)$, $S_{re}=Re[S(\textbf{p},t_s)]$ and $S_{im}=Im[S(\textbf{p},t_s)]$, and the dipole matrix element $\textbf{d}_i(\textbf{v})=\langle {\textbf{v}}| \textbf{r} |0\rangle$ \cite{Lewenstein1995}.  The corresponding complex amplitude $c'(\textbf{p}',t_i)$ for the Coulomb-modified trajectory $(\textbf{p}',t_i)$ in TRCM can be written as $c'(\textbf{p}',t_i)\propto { \textbf{E}(t_s) \cdot \textbf{d}_i[\textbf{p}+\textbf{A}(t_s)]e^{-i[S_{re}(\textbf{p}',t'_s)+iS_{im}(\textbf{p},t_s)]}}$ with $t'_s=t_i+it_x$. The above expression implies that $M'(\textbf{p}')=\left| c'(\textbf{p}',t_i) \right|^2=\left| c(\textbf{p},t_0) \right|^2=M(\textbf{p})$, allowing us to analytically decouple field-related effects related to the amplitude $M(\textbf{p})$ and Coulomb-related effects related to the shifted momentum $\textbf{p}'$ in ATI. 
	With the Coulomb-shifted trajectory $(\textbf{p}',t_i)$, the offset angle in the general attoclock related to the most probable route (MPR) in PMD can be defined as \cite{Che2021,Che2023}
	\begin{equation}
		\tan \theta =p_x'/p_y'\approx [v_x(t_0)-A_x (t_i)]/[v_y(t_0)-A_y (t_i)].		
	\end{equation}
	The angle defined above does not take into account the influence of the ATI ring and will be used to understand the offset angle in energy-resolved attoclock.

	\begin{figure}[t]
		\begin{center}
			\rotatebox{0}{\resizebox *{8.5cm}{6.8cm} {\includegraphics {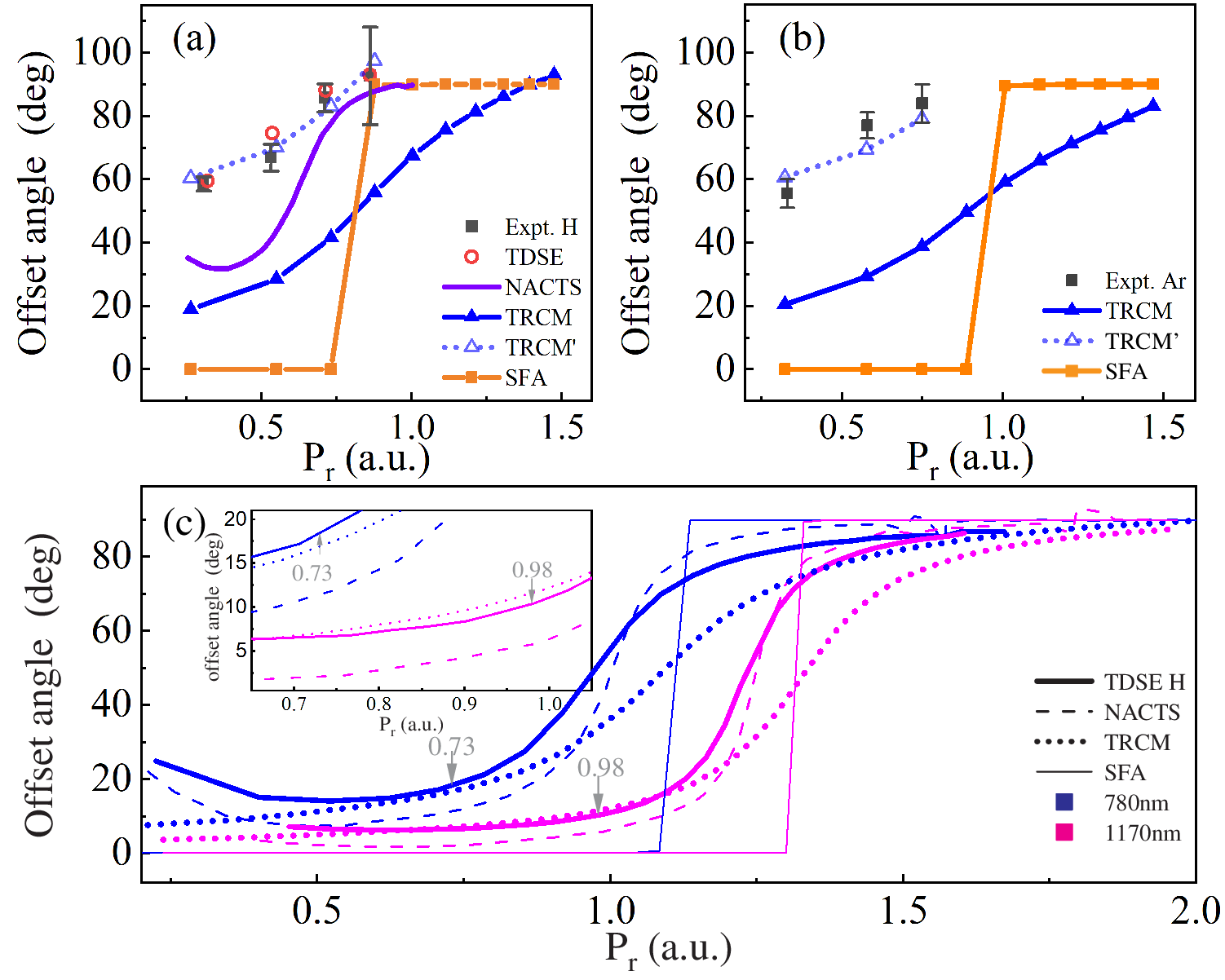}}}
		\end{center}
		\caption{Offset angles $\theta$ as a function of the radial momentum $p_r$ obtained by different methods for different targets and laser parameters. (a) The experimental, TDSE and model (NACTS) results for H from \cite{Trabert2021}. (b) The experimental results for Ar from \cite{Xie2024}.  (c) The TDSE and model results for H from \cite{Trabert2021}. In each panel, besides TRCM, results of SFA are also presented. In (a) and (b), the TRCM results are also vertically shifted to match the experimental data, represented by TRCM'. The laser parameters used in our calculations are $I=0.9\times10^{14}$W/cm$^2$, $\xi=0.85$ with $\lambda=390$ nm in (a) and $\lambda=780$ nm and $\lambda=1170$ nm in (c), and $I=0.9\times10^{14}$W/cm$^2$, $\xi=0.8$ with $\lambda=400$ nm in (b). }
		\label{fig1}
	\end{figure}
	
	\begin{figure}[t]
		\begin{center}
			\rotatebox{0}{\resizebox *{8.5cm}{11cm} {\includegraphics {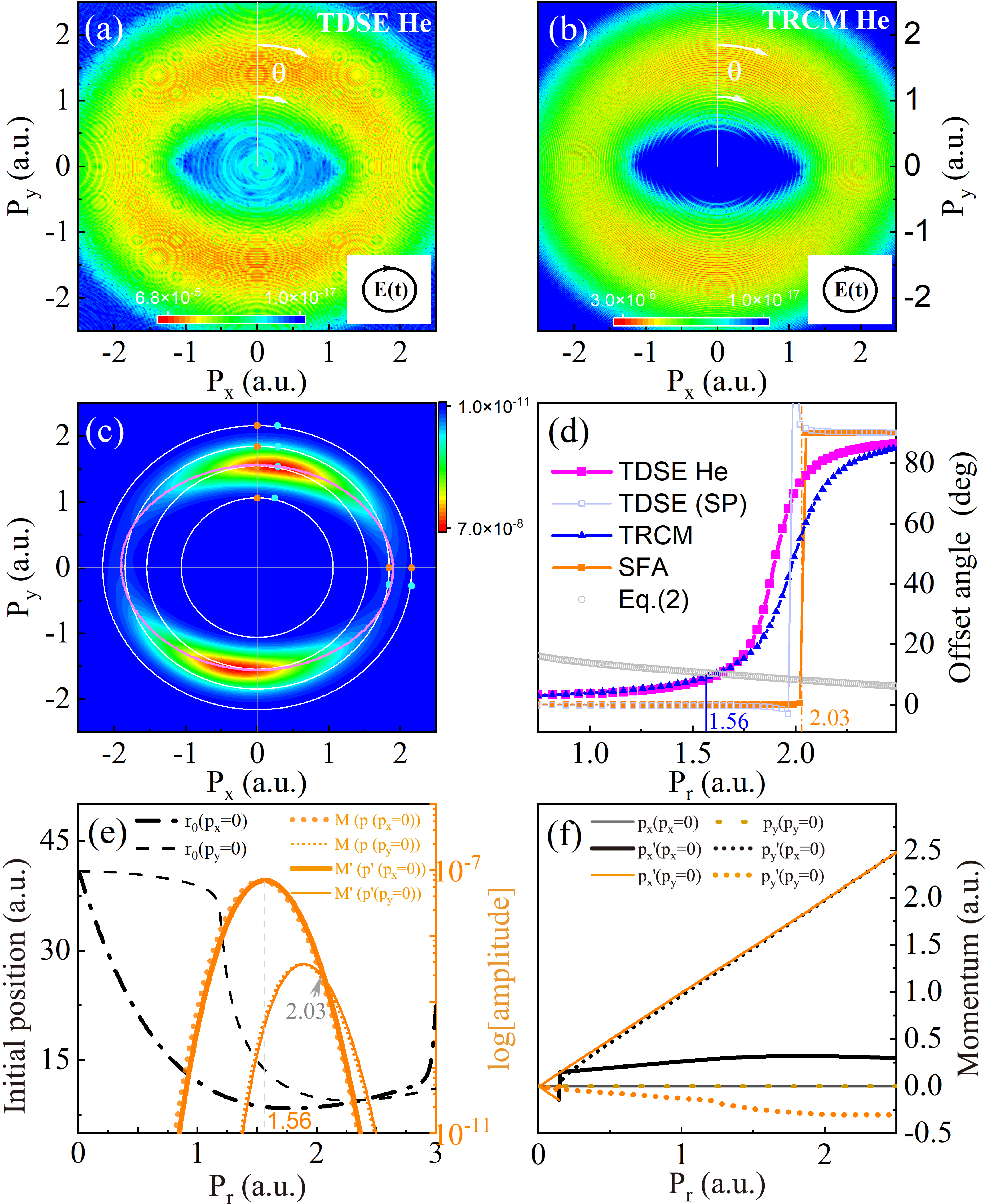}}}
		\end{center}
		\caption{Illustration of mechanism of energy-resolved attoclock. (a) and (b) PMDs of TDSE and TRCM for He. The inset shows the helicity of the EPL field and the orientation of the polarization ellipse. $\theta$ indicates the offset angle. (c) A sketch of ionization geometry. The color coding indicates the noncoherent PMD of TDSE in (a). The white concentric circles indicate some ATI rings of SFA predictions. The orange-solid dots indicate some typical momenta $\textbf{p}$ along the axes of $p_{x(y)}=0$ in the rings, denoted with $\textbf{p}(p_{x(y)}=0)$. The light-blue-solid dots indicate the corresponding Coulomb-shifted momenta $\textbf{p}'(p_{x(y)}=0)$ of Eq. (1). The pink dots indicate the LMPR abstracted from the noncoherent PMD of TRCM in (b) and show an elliptical structure. (d) Offset angles $\theta$ as a function of the radius momentum $p_r$, obtained from PMDs in (a) and (b). Offset angles from PMDs of SFA and TDSE with a short-range potential (SP) are also shown here. The gray-dot curve shows the results obtained by Eq. (2) for the Coulomb-shifted momenta $\textbf{p}'(p_{x}=0)$. (e) SFA predictions of initial positions $r_0(\textbf{p})$ and amplitudes $M(\textbf{p})$ for the momenta $\textbf{p}(p_{x(y)}=0)$. The corresponding TRCM predictions of amplitudes $M'(\textbf{p}')$ for $\textbf{p}'(p_{x(y)}=0)$ are also shown here. (f) Comparisons of $\textbf{p}(p_{x(y)}=0)$ and $\textbf{p}'(p_{x(y)}=0)$. The laser parameters are $I=5\times10^{14}$W/cm$^2$, $\lambda=800$ nm and $\xi=0.85$.}
		\label{fig2}
	\end{figure}

	\begin{figure}[t]
		\begin{center}
			\rotatebox{0}{\resizebox *{8.5cm}{8.5cm} {\includegraphics {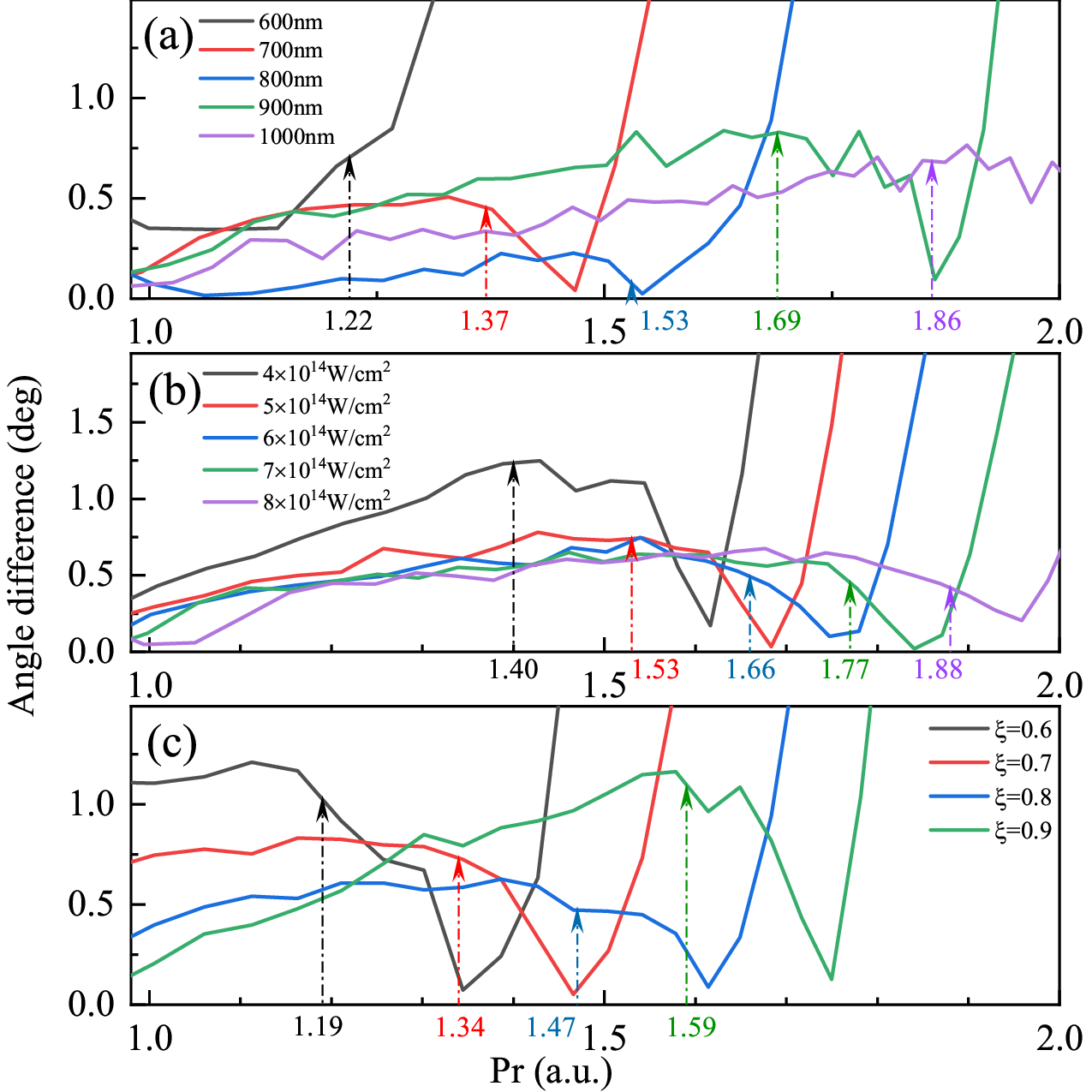}}}
		\end{center}
		\caption{Absolute differences of offset angles $\theta$ between TDSE and TRCM for He as a function of the radius momentum $p_r$, obtained for varied laser parameters.  (a) $I=5\times10^{14}$W/cm$^2$, $\xi=0.85$ with different $\lambda$. (b)  $\lambda=800$ nm, $\xi=0.85$ with different $I$. (c) $I=5\times10^{14}$W/cm$^2$, $\lambda=800$ nm with different $\xi$. The vertical arrows in (a-c) indicate the positions of MPR predicted by TRCM in each curve. }
		\label{fig3}
	\end{figure}
	
	\textit{Comparison to experiments}. In Fig. 1(a), we first present the comparison to the experimental offset angle of H \cite{Trabert2021}. The experimental angle shows a significant increasing trend with energy. The prediction of TRCM (blue-solid triangle) differs from the experimental data in quantity, but it well reproduces the scaling law of the data relative to energy. The vertically shifted TRCM results (blue-hollow triangle) agree well with these experimental data, suggesting that the TRCM holds the basic physics here. This quantitative difference may arise from the reason that the experimental data is obtained at a large Keldysh parameter \cite{Keldysh1965} of $\gamma\approx3$ at which the electron-trajectory theory and therefore the TRCM does not work very well. More insights are obtained from the SFA results (orange-solid square) which also show a sharp increasing trend for this angle, with $\theta=0^o$ for small energy and $\theta=90^o$ for high energy. In contrast, the experimental and TRCM results show a gradually increasing trend. In Fig. 1(b), we show the comparison to the experimental offset angles of Ar \cite{Xie2024}. In this case, the TRCM still reproduces the scaling law of the experimental data to energy. In addition, the sharp-increasing phenomenon is also observed in the SFA results here, suggesting that this phenomenon is inherent in SFA predictions and is mainly related to the properties of the laser field.  
	
	In Fig. 1(c), we show the comparison to three-dimensional TDSE results of H for cases of longer wavelengths obtained in \cite{Trabert2021}. For these cases with smaller Keldysh parameters $\gamma$, the predictions of TRCM for the offset angle become to quantitatively agree with the TDSE results, especially for the energy region around the MPR (indicated by the vertical arrows) where the PMD has large amplitudes. For example, for $\lambda=780$ nm with $\gamma=1.14$, the difference between TDSE and TRCM around the MPR of $p_r=0.73$ a.u. is about 1.5 degrees. For $\lambda=1170$ nm with $\gamma=0.75$, the difference around the MPR of $p_r=0.98$ a.u. is about 1.2 degree. The main difference between TDSE and TRCM lies in the energy region where the angle curve changes very quickly. This may be due to the intense competition mechanism between the contributions of the main and minor axes of the laser ellipse to ionization in the region, making quantitative description difficult, as to be discussed below.  
	
	\textit{Potential mechanism}. Based on the comparisons in Fig. 1, next, we discuss the potential mechanism. Since the increasing trend of the angle with energy is similar for short and long wavelengths and the TRCM works better for long wavelengths, in the following, we focus on the cases of long wavelength. To explore a wider parameter region, we also perform simulations with two-dimensional TDSE for He atoms. Relevant calculation details can be found in \cite{Che2023}. In Figs. 2(a) and 2(b), We show the PMDs of He obtained from TDSE and TRCM at $\gamma=0.64$. One can observe that in this case, the ATI rings can still be resolved here. Particularly, these rings have large amplitudes around the region of $1a.u.<p_r<1.74 a.u.$ (also see Fig. 2(c)). By using the numerical procedures of integration and Fourier transform as introduced in \cite{Trabert2021}, we can obtain the offset angles of each ring for both TDSE and TRCM results, as shown in Fig. 2(d). To illuminate the potential mechanism, in Fig. 2(c), we show the noncoherent PMD of TDSE and the local most probable route (LMPR)  \cite{Chao2021} from PMD of TRCM, which crosses the relatively bright part of the entire PMD of TDSE and presents an elliptical structure (pink dots). SFA predictions of some ATI rings (circular rings) and some possible local most probable momenta $(p_{x}=0,p_y)$ or $(p_{x},p_y=0)$ (orange-solid dots) in the rings, which are denoted with $\textbf{p}(p_{x(y)}=0)$, are also shown here. 
	
	According to Eq. (1), when the Coulomb effect is considered, these momenta $\textbf{p}(p_{x(y)}=0)$ are shifted to other momenta $\textbf{p}'(p_{x(y)}=0)$, which are denoted using the light-blue-solid dots. However, this shifts occurs along an elliptical orbit due to the property of the EPL field, as shown by the elliptical structure of LMPR (pink dots). Therefore, the shifted momenta generally deviate somewhat from the corresponding original ATI rings. This implies that the local most probable momenta $\textbf{p}_c$ found in each ATI ring of the Coulomb-included PMD generally are not the Coulomb-shifted ones of $\textbf{p}'(p_{x(y)}=0)$. However, around the MPR corresponding to the brightest part of the PMD, the distribution has large amplitudes. In this case, the momenta $\textbf{p}_c$ found from ATI rings may be near to the Coulomb-shifted ones. These analyses are supported by the angle results shown in Fig. 2(d). Indeed, in the energy region around the MPR of $p_r=1.56$ a.u., the energy-resolved angles are taken from the PMD of TRCM are very near to the TDSE ones with differences smaller than 1 degree (also see Fig. 3). In contrast, the angles predicted by Eq. (2) for the Coulomb-shifted momenta $\textbf{p}'(p_{x(y)}=0)$ (gray-hollow circles) deviate remarkably from the data obtained from the ATI rings in most of energy regions. However, in the energy region near the MPR of $p_r=1.56$ a.u., the predictions of Eq. (2) are near to these ring-related data. This observation also holds for other laser parameters in our simulations, suggesting that the ring-resolved offset angles near the MPR can be used to directly infer time delay information through Eq. (2). 
	
	In Fig. 2(d), we also show the results of TDSE with a short-range potential. The curve of short-range potential is very near to the SFA result, with angles jumping from $0^o$ to $90^o$ near $p_r=2.03$ a.u.. These typical features of the Coulomb-free curves are similar to the Coulomb-included ones on the whole, but the Coulomb-included curves are smooth. Therefore, we conclude that this trend of angle increasing with energy originates from the properties of the EPL field, and is softened by the Coulomb effect. 
	
	To analyze the field effects, in Fig. 2(e), we show the amplitudes and initial positions of SFA predictions for the typical momenta $\textbf{p}(p_{x(y)}=0)$,  which correspond to a zero-degree (90-degree) offset angle. The initial positions $r_0(\textbf{p})\equiv{r}_0(t_0)$ of these momenta both show a trend of decreasing first and then increasing with energy. The maximal amplitudes $M(\textbf{p})$ of these momenta along the axes of $p_x=0$ and $p_y=0$ basically correspond to the minimal exit positions $r_0(\textbf{p})$ of these momenta.  The maximal amplitude of $p_x=0$  arrives at the radius momentum of $p_r=1.56$ a.u. and the amplitude of $p_y=0$ (corresponding to $\theta=90^o$) becomes larger than that of $p_x=0$  ($\theta=0^o$)  at $p_r=2.03$ a.u.. The corresponding TRCM predictions of amplitudes $M'(\textbf{p}')$ for Coulomb-shifted momenta $\textbf{p}'(p_{x(y)}=0)$  deviate slightly from the SFA predictions.  These critical parameters of $p_r=1.56$ a.u. and $p_r=2.03$ a.u. explain the important characteristics of the curves discussed in Fig. 2(d). Finally, as a direct comparison, in Fig. 2(f), we plot the typical momenta $\textbf{p}(p_{x(y)}=0)$ and the Coulomb-shifted ones $\textbf{p}'(p_{x(y)}=0)$ of Eq. (1). Results of ${p_{y(x)}}(p_{x(y)}=0)$ (not shown here) are similar to ${p'_{y(x)}}(p_{x(y)}=0)$. These results further visualize the Coulomb effect and the origin of relevant theory curves in Fig. 2(d).   
	
	\textit{Extended comparisons}. To validate the above discussions, we further compare the absolute differences between predictions of TDSE and TRCM for He in a wide parameter region of laser intensity, wavelength and ellipticity. It can be observed from Figs. 3(a) to 3(c), around the MPR as indicated by the color vertical arrows, the differences are generally smaller than or near to 1 degree. Only for the case of relatively low intensity of $I=4\times10^{14}$W/cm$^2$ in Fig. 3(b), the difference is about 1.25 degrees, indicating the applicability of TRCM for energy-resolved attoclock in a wide parameter region. 
	
	\textit{Conclusion}. In summary, we have studied the energy-resolved attoclock, recently proposed for achieving high resolution in both time and energy, through numerical and analytical methods. The use of a recently developed Coulomb-included strong-field model, allows us to analytically and quantitatively decouple complex nonadiabatic and Coulomb effects, making the potential mechanism accessible. We have shown that the typical phenomenon of angle increasing with energy in energy-resolved attoclock can be attributed to the competing contributions of these two components of the EPL field to tunneling and the soft modulation of electron tunneling dynamics by the near-nuclear Coulomb potential. Since the model can give quantitative results at small Keldysh paramters and applicable scaling laws at large Keldysh parameters, it provides a promising analytical method for quantitatively studying ATI and other tunneling-triggered strong-field processes such as HHG. The combination of the energy-resolved attoclock and the method also provides the possibility for analytically and quantitatively studying nonadiabatic and Coulomb effects on tunneling dynamics of more complex systems such as aligned symmetric and polar molecules.

	This work was supported by the National Natural Science Foundation of China (Grant Nos. 12574376, 12404330, 12304303, 12174239).

\end{document}